\documentclass[aps,preprint,nofootinbib,eqsecnum]{revtex4}%
\usepackage{amsmath}
\usepackage{graphicx}
\usepackage{amsfonts}
\usepackage{amssymb}%
\providecommand{\U}[1]{\protect\rule{.1in}{.1in}}
\providecommand{\U}[1]{\protect\rule{.1in}{.1in}}
\providecommand{\U}[1]{\protect\rule{.1in}{.1in}}
\providecommand{\U}[1]{\protect\rule{.1in}{.1in}}
\providecommand{\U}[1]{\protect\rule{.1in}{.1in}}
\providecommand{\U}[1]{\protect\rule{.1in}{.1in}}
\providecommand{\U}[1]{\protect\rule{.1in}{.1in}}

\begin{document}
\preprint{{\leftline {USC-08/HEP-B3 \hfill }}}
\bigskip
\title[N=2,4 SUSY 2T-physics]{N=2,4 Supersymmetric Gauge Field Theory in 2T-physics}
\author{Itzhak Bars and Yueh-Cheng Kuo}
\affiliation{Department of Physics and Astronomy, University of Southern California, Los
Angeles, CA 90089-0484, USA\vspace{1cm}}
\thanks{This work was partially supported by the US Department of Energy, grant number DE-FG03-84ER40168.}

\begin{abstract}
In the context of Two Time Physics in 4+2 dimensions we construct the most general N=2,4 supersymmetric Yang Mills
gauge theories for any gauge group G. This builds on our previous work for N=1 supersymmetry. The action, the
conserved SUSY currents, and the off-shell SU(N) covariant SUSY transformation laws are presented for both N=2 and
N=4. The on-shell SUSY transformations close to the supergroup SU(2,2$|$N) with N=1,2,4. The SU(2,2)=SO(4,2)
sub-symmetry is realized linearly on 4+2 dimensional flat spacetime. All fields, including vectors and spinors,
are in 4+2 dimensions. The extra gauge symmetries in 2T field theory, together with the kinematic constraints that
follow from the action, remove all the ghosts to give a unitary theory. By choosing gauges and solving the
kinematic equations, the 2T field theory in 4+2 flat spacetime can be reduced to various shadows in various 3+1
dimensional (generally curved) spacetimes. These shadows are related to each other by dualities. The conformal
shadows of our theories in flat 3+1 dimensions coincide with the well known counterpart N=1,2,4 supersymmetric
massless renormalizable field theories in 3+1 dimensions. It is expected that our more symmetric new structures in
4+2 spacetime may be useful for non-perturbative or exact solutions of these theories.

\end{abstract}
\volumeyear{year}
\volumenumber{number}
\issuenumber{number}
\eid{identifier}
\date[]{}
\startpage{1}
\endpage{ }
\maketitle
\tableofcontents

\newpage

\section{Status of 2T-physics}

Two time physics (2T-physics) has proven to be successful in uniting different
ordinary 1T physics systems by establishing duality relationships among them
and in uncovering underlying hidden symmetries of 1T systems at the particle
and field theory levels \cite{2treviews}-\cite{2tGravity}. The theory starts
by imposing a Sp$(2,R)$ gauge symmetry on the phase space $\left(  X^{M}%
,P_{M}\right)  $ of a worldline theory of a bosonic particle \cite{2treviews}.
The local symmetry is generalized for spinning particles \cite{spin2t}%
\cite{spinBO}, supersymmetric particles \cite{super2t}\cite{2ttwistor}%
\cite{twistorBP}\cite{twistorLect}, or particles moving in background fields
\cite{2tbacgrounds}, but always involves Sp$\left(  2,R\right)  $ as a
subgroup. This symmetry requires that covariant momentum and position are
interchangeable at any instant for any motion. One finds that this symmetry
cannot exist in a spacetime with only one timelike dimension, and can be
realized without ghosts only in a spacetime with 2 timelike dimensions, no
less and no more.

It turns out that various usual 1T theories in $\left(  d-1\right)  +1$
dimensions are united by casting them into various gauge fixed versions of a
single parent 2T theory in $d+2$ dimensions. The relationship between the 1T
theories and the parent 2T theory is somewhat analogous to the relationship
between an object moving in a 3-dimensional room and the many shadows, with
their apparently unrelated motions, that can be created on walls by shining
light from different perspectives on the parent object. For example, the 1T
physics shadows created from the simplest 2T-physics bosonic particle that has
no parameters, include 1T particles with or without mass, moving in flat or
certain curved spacetimes, free or interacting in various potentials, and
their twistor equivalents. Some of the mathematical properties of these gauge
choices\footnote{For a graphical display of gauge choices see
http://physics.usc.edu/$\sim$bars/shadows.pdf} are summarized in three tables
in \cite{emergentfieldth1}. Through this procedure, a web of duality
relationships between these 1T theories with various parameters is established
as gauge transformations of the underlying 2T theory. This was most clearly
understood in the worldline formalism \cite{2treviews},\cite{2tHandAdS}%
-\cite{twistorBP}, and to some extent was also shown to be the property of 2T
field theory in $d+2$ dimensions \cite{emergentfieldth1}%
\cite{emergentfieldth2}. This is a new type of unification.

2T field theory is closely related to the underlying particle 2T worldline
theory by the BRST quantization procedure which, for the spinless particle,
followed a somewhat similar path \cite{2tbrst2006} to the BRST approach for
string field theory \cite{wittenSFT}. After integrating out redundant ghost
fields, this showed a simplified general way \cite{2tstandardM} to elevate 2T
worldline theories to the 2T field theory formalism. By now the Standard Model
and General Relativity have been shown to arise as particular shadows of their
respective parent 2T field theories for the Standard Model \cite{2tstandardM}
and for gravity \cite{2tGravity} in $d+2$ dimensions.

It was shown that the shadows derived from 2T field theory come with some
additional restrictions that are not present in the usual 1T field theory
approach. In particular, for the conformal shadow of the Standard Model mass
terms are not allowed. Then, in the Higgs scenario, the electroweak phase
transition needs to be driven by an additional scalar field which could be the
dilaton or another new SU$\left(  3\right)  \times$SU$\left(  2\right)
\times$U$\left(  1\right)  $ singlet scalar \cite{2tstandardM}\footnote{After
this proposal was discussed in \cite{2tstandardM} as part of possible new
physics signatures motivated by ideas in 2T-physics, similar scenarios that
include such a scalar field have been discussed in recent papers in both
theoretical and phenomenological contexts \cite{shapashnikov}-\cite{foot1}.}.
If it is the dilaton, this suggests that the Standard Model must be coupled to
the gravity sector in more ways than expected before. Given this, the
electroweak phase transition gets conceptually related to other phase
transitions that occurred in the history of the universe for which an
expectation value for the dilaton also plays a role.

Moreover, if ordinary General Relativity in $\left(  d-1\right)  +1$
dimensions is the conformal shadow of its 2T field theory counterpart in $d+2$
dimensions, then all scalar fields coupled to it must be conformal scalars
\cite{2tGravity}. This means that in addition to their usual coupling to the
spacetime metric $g_{\mu\nu}$, every scalar field $\phi\left(  x\right)  $
must also couple to the curvature scalar in the form $\left(  -a\phi
^{2}\right)  R\left(  g\right)  ,$ with the special unique coefficient
$a=\left(  d-2\right)  /8\left(  d-1\right)  $ in $d$ dimensions. In addition,
the gravitational constant arises only from the vacuum value of such scalars,
while a local Weyl symmetry removes a would be massless Goldstone dilaton.
This leads to new concepts in cosmology, including the possibility of a
changing gravitational constant as a result of various phase transitions in
the history of the Universe \cite{2tGravity}.

There is another interesting role for conformal scalars. It was suggested in
the second reference in \cite{2tstandardM} that a conformal scalar with its
required SO(4, 2) conformal symmetry could provide an alternative to
supersymmetry as a mechanism that could address the mass hierarchy problem.
This possibility has been more recently elaborated in \cite{nicolai2}%
\cite{foot2}.

It is remarkable that such new restrictions on 1T field theory that arise from
2T physics are compatible with current experimental knowledge and provide some
new conceptual and phenomenological guidance. Further developments on these
aspects will be reported elsewhere.

The 2T physics version of supersymmetric field theory in $4+2$ dimensions has also been developed by us in
previous papers \cite{susy2tN1}. In view of the remarks above, it is not surprising that the emergent
supersymmetric shadows also come with new corresponding constraints. Given the phenomenological interest in the
possibility of observing supersymmetry (SUSY) at the Large Hadron Collider (LHC), the 2T-physics constraints could
be phenomenologically significant, and we intend to study this topic in the near future.

In this paper we discuss the generalization of our previous work from $N=1$ to
$N=2$ and $N=4$ SUSY theories in $4+2$ dimensions. It should be noted that the
famous $N=4$ super Yang Mills gauge (SYM) theory in $3+1$ dimensions will
emerge as the conformal shadow of our $N=4$ SYM theory in $4+2$ dimensions.
Therefore, it will have other dual shadows which may be useful in the further
exact studies of this theory. The current paper will serve as a foundation for
later exploration of the structure and phenomena of these $N=1,2,4$ theories
and supersymmetric theories in higher dimensions.

\section{SO(4,2) spinors and notation \label{notation}}

In this section we briefly describe some of the notation used in this article.
For $N=2$ SUSY there are two left handed SO$\left(  4,2\right)  $ spinors
$\left(  \lambda_{L}\right)  _{i\alpha}^{a},$ where the label $i=1,2$
indicates the doublet of the SU$\left(  2\right)  $ R-symmetry, the label $a$
is for the adjoint representation of a compact gauge group $G$ and the label
$\alpha=1,2,3,4$ is for the $4$ representation of SU$\left(  2,2\right)  $
(left handed Weyl spinor of SO$\left(  4,2\right)  $). For the spinors
$\left(  \psi_{L}\right)  _{\alpha m}$ the label $m$ is used for some
arbitrary representation (including reducible representations) of the gauge
group $G$. Often we will simply use the label $L,$ suppressing the label
$\alpha$ to indicate a left handed spinor as $\lambda_{Li}^{a}$ or $\psi
_{Lm}.$ Sometimes we will also use the right handed spinor $\left(
\lambda_{R}\right)  _{\dot{\alpha}}^{ia},\left(  \psi_{R}\right)
_{\dot{\alpha}}^{m}$ in the $\bar{4}$ representation of SU$\left(  2,2\right)
,$ which is labeled with $\dot{\alpha}=1,2,3,4.$ One could rewrite all
right-handed spinors as left-handed ones by charge conjugation which is given
by%
\begin{equation}
\left(  \lambda_{R}\right)  _{a}^{i}\equiv\left(  C\overline{\lambda_{L}}%
^{T}\right)  _{a}^{i}=C\eta^{T}\left(  \lambda_{L}^{\ast}\right)  _{a}%
^{i},\;\;\text{or\ \ \ }\left(  \overline{\lambda_{L}}\right)  _{a}%
^{i}=-\left(  \lambda_{Ra}^{i}\right)  ^{T}C, \label{cc}%
\end{equation}
and similarly for $\psi.$ Here we have used the following matrices
\begin{equation}
\varepsilon=\left(
\genfrac{}{}{0pt}{}{0}{-1}%
\genfrac{}{}{0pt}{}{1}{0}%
\right)  ,\;C=\tau_{1}\times\sigma_{2},\;\eta=-i\tau_{1}\times1.
\end{equation}
where $\varepsilon_{ij}=-\varepsilon_{ji}$ is the antisymmetric charge
conjugation matrix for the SU$\left(  2\right)  $ R-symmetry, $C_{\dot{\alpha
}\beta}=-C_{\beta\dot{\alpha}}$ is the antisymmetric charge conjugation matrix
in SU$\left(  2,2\right)  $ spinor space, and $\eta^{\dot{\alpha}\beta}%
=\eta^{\beta\dot{\alpha}}$ is the symmetric SU$\left(  2,2\right)  $ metric in
spinor space used to construct the SU$\left(  2,2\right)  $ contravariant
spinor from the Hermitian conjugate spinor
\begin{equation}
\left(  \overline{\lambda_{L}}\right)  _{a}^{i\beta}=\left(  \left(
\lambda_{iL}^{a}\right)  ^{\dagger}\eta\right)  ^{\beta}=\left(  \lambda
_{L}^{\dagger}\right)  _{\dot{\alpha}a}^{i}\eta^{\dot{\alpha}\beta}.
\end{equation}
Note that Hermitian conjugation $\left(  \psi_{L\alpha m}\right)  ^{\dagger
}=\left(  \psi_{L}^{\dagger}\right)  _{\dot{\alpha}}^{m}$ changes the
SU$\left(  2,2\right)  $ index from $\alpha$ to $\dot{\alpha}$ and raises the
index $m$ assuming $m$ labels a complex representation of the gauge group $G.$
Similarly, $\left(  \lambda_{Li\alpha}^{a}\right)  ^{\dagger}=\left(
\lambda_{L}^{\dagger}\right)  _{a\dot{\alpha}}^{i}$ raises the SU$\left(
2\right)  $ index $i$ and drops the index $a.$ However, the adjoint
representation is real, the Killing metric $\delta_{ab}$ can be taken as $1,$
so that there is no distinction between upper and lower $a$ indices, and the
structure constants $f_{abc}$ are completely antisymmetric. Using these
definitions we can also write the following relations that are equivalent to
(\ref{cc})
\begin{equation}
\lambda_{Li}^{a}=-\left(  C\overline{\lambda_{aR}}^{T}\right)  _{i}\text{,
\ ~or~~\ }\left(  \overline{\lambda_{R}}\right)  _{i}^{a}=\left(  \lambda
_{Li}^{a}\right)  ^{T}C. \label{cc2}%
\end{equation}
The SU$\left(  2\right)  $ indices $i$ may be further dropped or raised by
using the antisymmetric $\varepsilon$ and its inverse $\varepsilon
^{-1}=-\varepsilon$ as follows, $\lambda_{i}=\varepsilon_{ij}\lambda^{j}$ and
$\lambda^{i}=-\varepsilon^{ij}\lambda_{j}.$

We use the following explicit form of $4\times4$ SO$\left(  4,2\right)  $
gamma matrices $\Gamma^{M},\bar{\Gamma}^{M}$ in the Weyl bases ($M=0^{\prime
},1^{\prime},0,1,2,3$ is the label for the vector of SO$\left(  4,2\right)
$)
\begin{align}
\Gamma^{0^{\prime}}  &  =-i\tau_{1}\times1,\;\;\Gamma^{1^{\prime}}=\tau
_{2}\times1,\;\;\Gamma^{0}=1\times1,\ \;\;\;\Gamma^{i}=\tau_{3}\times
\sigma^{i}\label{gam1}\\
\bar{\Gamma}^{0^{\prime}}  &  =-i\tau_{1}\times1,\;\;\bar{\Gamma}^{1^{\prime}%
}=\tau_{2}\times1,\;\;\bar{\Gamma}^{0}=-1\times1,\ \;\;\bar{\Gamma}^{i}%
=\tau_{3}\times\sigma^{i} \label{gam2}%
\end{align}
These are compatible with the metric $\eta$ and the charge conjugation matrix
$C$ given above as explained in detail in Appendix (A) of ref.(\cite{susy2tN1}%
). In particular we note the hermiticity and charge conjugation properties
\begin{equation}%
\begin{array}
[c]{ll}%
\eta\Gamma^{M}\eta^{-1}=-\left(  \bar{\Gamma}^{M}\right)  ^{\dagger},\;\; &
\eta\bar{\Gamma}^{M}\eta^{-1}=-\left(  \Gamma^{M}\right)  ^{\dagger},\\
C\Gamma^{M}C^{-1}=\left(  \Gamma^{M}\right)  ^{T},\;\; & C\bar{\Gamma}%
^{M}C^{-1}=\left(  \bar{\Gamma}^{M}\right)  ^{T}.\;\;
\end{array}
\end{equation}
The matrices $\left(  \Gamma^{MN}\right)  _{\alpha\dot{\beta}}\equiv\frac
{1}{2}\left(  \Gamma^{M}\bar{\Gamma}^{N}-\Gamma^{N}\bar{\Gamma}^{M}\right)
_{\alpha\dot{\beta}}$ and $\left(  \Gamma^{M}\bar{\Gamma}^{N}+\Gamma^{N}%
\bar{\Gamma}^{M}\right)  _{\alpha\dot{\beta}}=2\delta_{\alpha\dot{\beta}}%
\eta^{MN},$ together with the antisymmetric matrices $\left(  \Gamma
^{M}C\right)  _{\alpha\beta}$,$\left(  C\bar{\Gamma}\right)  _{\dot{\alpha
}\dot{\beta}}$ incorporate the group theoretical properties of SU$\left(
2,2\right)  =$SO$\left(  4,2\right)  $ products of representations
\begin{equation}
\left(  4\times\bar{4}\right)  =15+1,\;\;\left(  4\times4\right)
_{antisymmetric}=6,\text{ \ \ }\left(  \bar{4}\times\bar{4}\right)
_{antisymmetric}=6. \label{su22rule}%
\end{equation}

The matrix representation of the generators of the gauge group $G$ are denoted
as $\left(  t_{a}\right)  _{m}^{~n}$ implying the group transformation law
$\delta_{\omega}\varphi_{m}=-i\omega^{a}\left(  t_{a}\right)  _{m}^{~n}%
\varphi_{n}.$ For the adjoint representation $\left(  t_{a}\right)  _{m}^{~n}$
is replaced by $\left(  t_{a}\right)  _{b}^{~c}=-if_{ab}^{~~c}=-if_{abc}.$ The
matrices $\left(  t_{a}\right)  _{m}^{~n}$ or $\left(  t_{a}\right)  _{b}%
^{~c}$ satisfy the Lie algebra $\left[  t_{a},t_{b}\right]  =if_{ab}%
^{~~c}\left(  t_{c}\right)  .$

\section{N=2 SUSY from N=1 in 4+2 dimensions}

The starting point is the general $N=1$ supersymmetric Yang-Mills 2T field
theory in $4+2$ dimensions for any compact Yang-Mills gauge group $G$
\cite{susy2tN1}. The theory contains a single $N=1$ vector supermultiplet
$\left(  A_{M},\lambda_{L},B\right)  ^{a},$ where $a$ labels the adjoint
representation of $G,$ plus any number of $N=1$ chiral supermultiplets
$\left(  \varphi,\psi_{L},F\right)  _{r}$ where $r$ labels an arbitrary
representation of the gauge group $G.$ This representation can be taken to be
reducible, hence it may contain any number of chiral multiplets in various
irreducible representations of $G$.

The action consistent with both $N=1$ SUSY and 2T field theory was given in
\cite{susy2tN1} as follows%
\begin{equation}
S_{N=1}=\int d^{4+2}X~\delta\left(  X^{2}\right)  ~\left(  L_{kinetic}%
^{N=1}+L_{yukawa}^{N=1}+L_{potential}^{N=1}\right)  \label{N1action}%
\end{equation}
We note the typical delta function $\delta\left(  X^{2}\right)  ~$in 2T field
theory\footnote{The term $\frac{1}{2}\delta\left(  X^{2}\right)  \left(
\varphi^{a\dagger}D^{M}D_{M}\varphi_{a}+h.c.\right)  $ can also be written as
$-\delta\left(  X^{2}\right)  D^{M}\varphi^{a\dagger}D_{M}\varphi_{a}%
+2~\delta^{\prime}\left(  X^{2}\right)  ~\varphi^{a\dagger}\varphi_{a}$ after
an integration by parts, as in \cite{susy2tN1}.}, with a Lagrangian density
given by\footnote{The distinctive spacetime features including the delta
function $\delta\left(  X^{2}\right)  $ and its derivative that impose
$X^{M}X_{M}=0,$ as well as the explicit insertions of $X^{M}$ in the form of
$X=X_{M}\Gamma^{M}$ in the fermion kinetic terms and Yukawa couplings, are
required by the group theory rules of the spacetime SO$\left(  4,2\right)
=$SU$\left(  2,2\right)  $ in Eq.(\ref{su22rule}) and by the gauge symmetries
of 2T-physics field theory as explained in \cite{2tstandardM}.}
\begin{equation}
L_{kinetic}^{N=1}=\left\{
\begin{array}
[c]{l}%
-\frac{1}{4}F_{MN}^{a}F_{a}^{MN}+\frac{1}{2}\varphi^{r\dagger}D_{M}%
D^{M}\varphi_{r}+\frac{1}{2}\varphi_{r}D_{M}D^{M}\varphi^{r\dagger}\\
+\frac{i}{2}\left[  \overline{\lambda_{L}}^{a}X\bar{D}\lambda_{aL}%
+\overline{\lambda_{L}}^{a}\overleftarrow{D}\bar{X}\lambda_{aL}\right]
+\frac{i}{2}\left(  \overline{\psi_{L}}^{r}X\bar{D}\psi_{rL}+\overline
{\psi_{L}}^{r}\overleftarrow{D}\bar{X}\psi_{rL}\right)
\end{array}
\right\}  \label{n11}%
\end{equation}%
\begin{equation}
L_{yukawa}^{N=1}=\left[  \sqrt{2}g\varphi^{\dagger r}\left(  t_{a}\right)
_{r}^{~s}\left(  \psi_{sL}\right)  ^{T}\left(  C\bar{X}\right)  \lambda
_{L}^{a}-\frac{i}{2}\psi_{rL}\left(  C\bar{X}\right)  \psi_{sL}\frac
{\partial^{2}W}{\partial\varphi_{r}\partial\varphi_{s}}\right]
+~h.c.\label{n12}%
\end{equation}%
\begin{equation}
L_{potential}^{N=1}=\frac{1}{2}B^{a}B_{a}+F^{\dagger r}F_{r}+g\varphi^{\dagger
r}\left(  t_{a}\right)  _{r}^{~s}\varphi_{s}B^{a}+\left[  \frac{\partial
W}{\partial\varphi_{r}}F_{r}+h.c.\right]  \label{n13}%
\end{equation}
where $X\equiv X^{M}\Gamma_{M}$ and $\bar{D}\equiv\bar{\Gamma}^{M}D_{M}.$
These structures are compatible with the spacetime SU$\left(  2,2\right)
=$SO$\left(  4,2\right)  $ group theoretical rules in Eq.(\ref{su22rule}). The
explicit $X^{M}$ that appears in the kinetic, Yukawa, and the $\delta\left(
X^{2}\right)  ,$ is to be noted; hence there is no translation symmetry in 4+2
dimensions. However, the rotation symmetry SO$\left(  4,2\right)  $ turns into
conformal symmetry for the conformal shadow in $3+1$ dimensions, which
includes translation symmetry for the shadow in $3+1$ dimensions.

The superpotential $W$ is \textit{purely cubic}\footnote{The purely cubic
property of the superpotential is imposed by the 2T gauge symmetry
\cite{2tstandardM}. This implies that there are no dimensional parameters,
such as masses, in the potential. To induce mass terms in a nontrivial vacuum,
the dilaton must also be coupled to the other scalars as described in
\cite{2tstandardM} and in \cite{2tGravity}. This implies that the entire
supergravity multiplet, which includes the dilaton must also be included as
part of the theory of mass generation in the supersymmetric theory.
\label{cubic}} in the fields and is also $G$ invariant $\delta_{\omega
}W=-i\omega^{a}\frac{\partial W}{\partial\varphi_{r}}\left(  t_{a}%
\varphi\right)  _{r}=0.$ The field equations may be solved for the auxiliary
fields,
\begin{equation}
B_{a}=-g\varphi^{\dagger}t_{a}\varphi,\;F_{r}=-\frac{\partial W^{\dagger}%
}{\partial\varphi^{\dagger r}},\;F^{\dagger r}=-\frac{\partial W}%
{\partial\varphi_{r}}, \label{BF}%
\end{equation}
so that this theory contains just the fields $\left(  A_{M},\lambda
_{L}\right)  ^{a}$ and $\left(  \varphi,\psi_{L}\right)  _{r}.$ In
\cite{susy2tN1} it was demonstrated that this theory has $N=1$ supersymmetry,
and the corresponding conserved current in $4+2$ dimensions was given (see
also below).

To construct the general theory with $N=2$ supersymmetry in 2T field theory we follow the same strategy employed
in 1T SUSY field theory but modified to conform to 2T field theory structures\footnote{The method used here in
$4+2$ dimensions parallels a similar discussion in usual SUSY field theory in $3+1$ dimensions as described in
\cite{weinberg}. Note however that the explicit SUSY transformations in $4+2$ dimensions have many features that
are absent in the corresponding SUSY transformations in $3+1$ dimensions. Nevertheless those details do not play a
role in this method.}. We start with the general $N=1$ theory given above, with one $N=1$ vector multiplet, and 3
distinct representations of $N=1$ chiral multiplets embedded in the reducible representation labeled by $r$.
Namely, we consider the following $N=1$
supermultiplets%
\begin{align}
\text{vector }  &  \text{:~}\left(  A_{M},\lambda_{L}\right)  ^{a}%
,\;\text{chiral-0:\ }\left(  \varphi,\psi_{L}\right)  ^{a},\;\text{both in the
adjoint representation,}\\
\text{chiral-1}  &  \text{:\ }\left(  \phi,\eta_{L}\right)  _{n}%
,\;\text{chiral-2:\ }(\tilde{\phi},\tilde{\eta}_{L})^{n},\;\text{in arbitrary
complex conjugate repr.}%
\end{align}
So, the label $r$ in Eqs.(\ref{n11}-\ref{BF}) is now specialized to the 3
representations labeled by the adjoint $a$, lower $n$ and upper $n.$ The
reducible matrix representation $\left(  t_{a}\right)  _{r}^{~s}$ is also
specialized as follows
\begin{equation}
\left(  t_{a}\right)  _{r}^{~s}:\;\left(  t_{a}\right)  _{b}^{~c}%
=-if_{ab}^{~~~c},\;\text{and \ }\left(  t_{a}\right)  _{n}^{~m},
\end{equation}
implying the $G$ transformation rules
\begin{equation}
\delta_{\omega}\varphi_{b}=-\omega^{a}f_{ab}^{~~~c}\varphi_{c},\;\;\;\delta
_{\omega}\phi_{n}=-i\omega^{a}\left(  t_{a}\right)  _{n}^{~m}\phi
_{m},\;\;\;\delta_{\omega}\tilde{\phi}^{n}=i\omega^{a}\tilde{\phi}^{m}\left(
t_{a}\right)  _{m}^{~n}.
\end{equation}
The complex conjugate representations labeled by lower $n$ and upper $n$ can
themselves be reducible representations. In any case, $\tilde{\phi}^{m}%
\phi_{m}$ is invariant, while $\tilde{\phi}t_{a}\phi$ transforms like the
adjoint representation.

When the superpotential $W$ is taken of the following form
\begin{equation}
W=i\sqrt{2}g\tilde{\phi}t_{a}\phi\varphi^{a},\;
\end{equation}
there is automatically $N=2$ supersymmetry as well as local gauge symmetry
under the Yang-Mills group $G.$ To show the $N=2$ structure one writes the
$N=1$ Lagrangian following the recipe given above in Eqs.(\ref{n11}-\ref{BF}).
Then one can notice that there is a symmetry under the following discrete
transformation $[\lambda_{L}^{a}\rightarrow\psi_{L}^{a},\;\psi_{L}%
^{a}\rightarrow-\lambda_{L}^{a}]$ and $[\phi_{n}\rightarrow\tilde{\phi}%
_{n}^{\dagger},\;\tilde{\phi}_{n}^{\dagger}\rightarrow-\phi_{n}],$ or
equivalently $[\phi^{\dagger n}\rightarrow\tilde{\phi}^{n},\;\tilde{\phi}%
^{n}\rightarrow-\phi^{\dagger n}],$ while the other fields $A_{M}^{a}%
,\varphi^{a},\eta_{Ln},\tilde{\eta}_{L}^{n}$ remain unchanged. This
transformation is just a discrete subgroup of the SU$\left(  2\right)  $
$R$-symmetry which acts on the SU$\left(  2\right)  $ doublets $\left(
\genfrac{}{}{0pt}{}{\lambda_{L}^{a}}{\psi_{L}^{a}}%
\right)  ,\left(
\genfrac{}{}{0pt}{}{\phi_{n}}{\tilde{\phi}_{n}^{\dagger}}%
\right)  ,\left(
\genfrac{}{}{0pt}{}{-\tilde{\phi}^{n}}{\phi^{\dagger n}}%
\right)  $ as follows\footnote{The discrete transformation $R=e^{i\sigma
_{2}\pi/2}=i\sigma_{2}=\left(
\genfrac{}{}{0pt}{}{0}{-1}%
\genfrac{}{}{0pt}{}{1}{0}%
\right)  $ corresponds to a SU$\left(  2\right)  $ rotation by an angle $\pi
.$}
\begin{equation}
\left(
\genfrac{}{}{0pt}{}{\lambda_{L}^{a}}{\psi_{L}^{a}}%
\right)  ^{\prime}=\left(
\genfrac{}{}{0pt}{}{0}{-1}%
\genfrac{}{}{0pt}{}{1}{0}%
\right)  \left(
\genfrac{}{}{0pt}{}{\lambda_{L}^{a}}{\psi_{L}^{a}}%
\right)  =\left(
\genfrac{}{}{0pt}{}{\psi_{L}^{a}}{-\lambda_{L}^{a}}%
\right)  ;\;\;\;\left(
\genfrac{}{}{0pt}{}{\phi_{n}}{\tilde{\phi}_{n}^{\dagger}}%
\right)  ^{\prime}=\left(
\genfrac{}{}{0pt}{}{0}{-1}%
\genfrac{}{}{0pt}{}{1}{0}%
\right)  \left(
\genfrac{}{}{0pt}{}{\phi_{n}}{\tilde{\phi}_{n}^{\dagger}}%
\right)  =\left(
\genfrac{}{}{0pt}{}{\tilde{\phi}_{n}^{\dagger}}{-\phi_{n}}%
\right)  \label{Rdiscrete}%
\end{equation}
The last relation can also be written equivalently for the charge conjugate
doublet $\left(
\genfrac{}{}{0pt}{}{-\tilde{\phi}^{n}}{\phi^{\dagger n}}%
\right)  ^{\prime}=\left(
\genfrac{}{}{0pt}{}{0}{-1}%
\genfrac{}{}{0pt}{}{1}{0}%
\right)  \left(
\genfrac{}{}{0pt}{}{-\tilde{\phi}^{n}}{\phi^{\dagger n}}%
\right)  .$ Actually, this Lagrangian has a global symmetry under the
continuous SU$\left(  2\right)  $ R-symmetry transformations applied on the
doublets above as will be made manifest in the next section.

Now we concentrate on identifying the second supersymmetry by starting with
the known \cite{susy2tN1} $N=1$ SUSY transformations of our fields%
\begin{equation}
\delta_{\varepsilon_{1}}(A_{M},\lambda_{L})^{a},\;\delta_{\varepsilon_{1}%
}\left(  \varphi,\psi_{L}\right)  ^{a},\;\delta_{\varepsilon_{1}}(\phi
,\eta_{L})_{n},\;\delta_{\varepsilon_{1}}(\tilde{\phi},\tilde{\eta}_{L})^{n}.
\label{susy1}%
\end{equation}
The expressions for these are given in \cite{susy2tN1} but for now we will not need them explicitly. It suffices
to know that the action above is invariant under this first SUSY transformation $\delta_{\varepsilon_{1}}$ with
parameter $\varepsilon_{1L},$ which is a left-handed chiral spinor labeled by $L=[$4 of SU$\left(  2,2\right)  ]$
\cite{susy2tN1}. Corresponding to this symmetry there is a conserved supercurrent in $4+2$ dimensions
$J_{1L}^{M},$ that satisfies $\partial_{M}J_{1L}^{M}=0$ when the equations of motion are used (see below).

Since we have already identified in Eq.(\ref{Rdiscrete}) a discrete $R$
symmetry of the Lagrangian, it must be that the action is invariant also under
a second SUSY transformation $\delta_{\varepsilon_{2}}$ with parameter
$\varepsilon_{2L}.$ The $\delta_{\varepsilon_{2}}$ transformation laws must
look the same as those of $\delta_{\varepsilon_{1}}$ after applying the
discrete transformation of Eq.(\ref{Rdiscrete}) on the expressions in
Eq.(\ref{susy1}) and then replacing $\varepsilon_{1L}$ by $\varepsilon_{2L}.$
Hence the second SUSY transformation is obtained from the first one as follows%
\begin{equation}
\delta_{\varepsilon_{2}}(A_{M},\psi_{L})^{a},\;\delta_{\varepsilon_{2}}\left(
\varphi,-\lambda_{L}\right)  ^{a},\;\delta_{\varepsilon_{2}}(\tilde{\phi
}^{\dagger},\eta_{L})_{n},\;\delta_{\varepsilon_{2}}(-\phi^{\dagger n}%
,\tilde{\eta}_{L})^{n}. \label{susy2}%
\end{equation}
We see that the second SUSY transformation $\delta_{\varepsilon_{2}}$ looks
like again a $N=1$ transformation, but the fields have been reshuffled into
new $N=1$ vector and chiral multiplets as seen by comparing Eqs.(\ref{susy1}%
,\ref{susy2}). For example the $\delta_{\varepsilon_{2}}$ SUSY partner of
$A_{M}$ is now $\psi_{L}$ rather than $\lambda_{L},$ and so on. With the same
discrete $R$ transformation technique applied on the expression for the
supercurrent $J_{1L}^{M}$ we can construct the second conserved SUSY current
$J_{2L}^{M}$ (see below).

\subsection{ SU$\left(  2\right)  $ covariant N=2 SUSY in 4+2 dims}

It is evident from the previous section that, once the discrete $R$ symmetry
has been identified, it is guaranteed that the theory has $N=2$ supersymmetry.
It is useful to make this SU$\left(  2\right)  $ and $N=2$ symmetry manifest
by using fields with SU$\left(  2\right)  $ doublet and singlet representation
labels, and then rewrite the action, conserved currents, and transformation
laws, described above in terms of these SU$\left(  2\right)  $
representations. The result is the following.

The doublets are labeled by an index $i=1,2$ as follows%
\begin{equation}
\lambda_{iL}^{a}=\left(
\genfrac{}{}{0pt}{}{\lambda_{1L}^{a}}{\lambda_{2L}^{a}}%
\right)  \equiv\left(
\genfrac{}{}{0pt}{}{\lambda_{L}^{a}}{\psi_{L}^{a}}%
\right)  ;\;\;\phi_{in}=\left(
\genfrac{}{}{0pt}{}{\phi_{1n}}{\phi_{2n}}%
\right)  \equiv\left(
\genfrac{}{}{0pt}{}{\phi_{n}}{\tilde{\phi}_{n}^{\dagger}}%
\right)  ,\;\;\varepsilon_{iL}=\left(
\genfrac{}{}{0pt}{}{\varepsilon_{1L}}{\varepsilon_{2L}}%
\right)
\end{equation}
while the other fields $A_{M}^{a},\varphi^{a},\eta_{Ln},\tilde{\eta}_{L}^{n}$
are SU$\left(  2\right)  $ singlets. It is also useful to introduce auxiliary
fields $S_{ij}^{a}$ and $F_{in},$ where $F_{in}$ is an SU$\left(  2\right)  $
doublet while $S_{ij}^{a}$ is a symmetric tensor representing a triplet of
SU$\left(  2\right)  $. It is convenient to collect these into one $N=2$
vector multiplet in the adjoint representation of $G$ and many $N=2$
hypermultiplets labeled by $n$ in some (generally reducible) representation
of $G$%
\begin{equation}
\text{vector :\ }\left(  A_{M}^{a},\lambda_{iL}^{a},\varphi^{a},S_{ij}%
^{a}\right)  ;\;\;\text{hyper : }\left(  \phi_{in},\eta_{nL},\tilde{\eta}%
_{nR},F_{in}\right)  ,\;i=1,2.
\end{equation}
Here we have used the charge conjugate right handed spinor $\tilde{\eta}%
_{Rn}=C\overline{\tilde{\eta}_{L}^{n}}^{T}$ instead of the original left
handed $\tilde{\eta}_{L}^{n}.$ In fact, the two SO$\left(  4,2\right)  $ Weyl
spinors $\left(  \eta_{Ln},\tilde{\eta}_{Rn}\right)  ,$ transforming as
$\left(  4\oplus\bar{4}\right)  $ of SU$\left(  2,2\right)  ,$ taken together
can be considered as a full 8 dimensional Dirac spinor of SO$\left(
4,2\right)  .$In what follows, we choose to present the theory without the
auxiliary fields.

The manifestly SU$\left(  2\right)  $ invariant $N=2$ action is%
\begin{equation}
S_{N=2}=\int d^{4+2}X~\delta\left(  X^{2}\right)  ~\left(  L_{kin}%
^{N=2}+L_{yukawa}^{N=2}+L_{potential}^{N=2}\right)  .
\end{equation}
The kinetic term is%
\[
L_{kin}^{N=2}=\left\{
\begin{array}
[c]{l}%
-\frac{1}{4}F_{MN}^{a}F_{a}^{MN}+\frac{i}{2}\left[  \overline{\lambda_{L}%
}^{ai}X\bar{D}\lambda_{iL}^{a}+\overline{\lambda_{L}}^{ai}\overleftarrow
{D}\bar{X}\lambda_{iL}^{a}\right] \\
+\frac{1}{2}\varphi^{\dagger a}D^{M}D_{M}\varphi^{a}+\frac{1}{2}\varphi
^{a}D^{M}D_{M}\varphi^{\dagger a}\\
+\frac{1}{2}\phi^{\dagger in}D_{M}D^{M}\phi_{in}+\frac{1}{2}\phi_{in}%
D_{M}D^{M}\phi^{in\dagger}\\
+\frac{i}{2}\left[  \overline{\eta_{L}}^{n}X\bar{D}\eta_{nL}+\overline
{\eta_{L}}^{n}\overleftarrow{D}\bar{X}\eta_{nL}\right] \\
-\frac{i}{2}\left[  \overline{\tilde{\eta}_{R}}^{n}\bar{X}D\tilde{\eta}%
_{nR}+\overline{\tilde{\eta}_{R}}^{n}\overleftarrow{\bar{D}}X\tilde{\eta}%
_{nR}\right]  \;
\end{array}
\right\}
\]
The Yukawa interactions are%
\[
L_{yukawa}^{N=2}=\left\{
\begin{array}
[c]{l}%
\sqrt{2}g\left(  t_{a}\right)  _{n}^{~m}\phi_{im}\left(  \varepsilon
^{ij}\overline{\tilde{\eta}_{R}}^{n}\bar{X}\lambda_{jL}^{a}+\overline{\eta
_{L}}^{n}X\lambda_{R}^{ai}\right) \\
+\frac{ig}{\sqrt{2}}\varepsilon^{ij}f_{abc}\varphi^{\dagger a}\lambda_{iL}%
^{b}C\bar{X}\lambda_{jL}^{c}+\sqrt{2}g\varphi^{a}\overline{\tilde{\eta}_{R}%
}\bar{X}t_{a}\eta_{L}%
\end{array}
\right\}  +h.c.
\]
The scalar potential term is%
\begin{equation}
L_{potential}^{N=2}=\left\{
\begin{array}
[c]{l}%
-g^{2}\left[  \phi^{\dagger i}t_{a}t_{b}\phi_{i}\right]  \left(  \varphi
^{a}\varphi^{\dagger b}+\varphi^{b}\varphi^{\dagger a}\right) \\
-g^{2}\phi_{(i}^{\dagger}t_{a}\phi_{j)}\phi^{\dagger(i}t_{a}\phi^{j)}-\frac
{1}{2}g^{2}\left(  if_{abc}\varphi^{\dagger b}\varphi^{c}\right)  ^{2}%
\end{array}
\right\}
\end{equation}
where $\phi^{\dagger(i}t_{a}\phi^{j)}\equiv\frac{1}{2}\left(  \varepsilon
^{jk}\phi^{\dagger i}t_{a}\phi_{k}+\varepsilon^{ik}\phi^{\dagger j}t_{a}%
\phi_{k}\right)  $.

If one desires, one could include auxiliary fields into the action by
introducing quadratic terms $\frac{1}{2}S_{a}^{ij}S_{ij}^{a}$ and $F^{\dagger
in}F_{in}$ and replacing the potential terms by%
\begin{equation}
L_{potential}^{N=2}=\left\{
\begin{array}
[c]{l}%
\frac{1}{2}S_{a}^{ij}S_{ij}^{a}+F^{\dagger in}F_{in}-g^{2}\left[
\phi^{\dagger i}t_{a}t_{b}\phi_{i}\right]  \left(  \varphi^{a}\varphi^{\dagger
b}+\varphi^{b}\varphi^{\dagger a}\right) \\
+\sqrt{2}g\left(  S^{a}\right)  _{i}^{~j}\phi^{\dagger i}t_{a}\phi_{j}%
-\frac{1}{2}g^{2}\left(  if_{abc}\varphi^{\dagger b}\varphi^{c}\right)  ^{2}%
\end{array}
\right\}
\end{equation}

The $N=2$ supercurrent is (in this expression $\lambda_{R}^{aj}\equiv\left(
C\overline{\lambda_{L}}^{ja}\right)  ^{T},$ $\eta_{R}^{n}\equiv\left(
C\overline{\eta_{L}}^{a}\right)  ^{T},\;\tilde{\eta}_{L}^{m}\equiv-\left(
C\overline{\tilde{\eta}_{R}}^{a}\right)  ^{T}$)%

\begin{equation}
J_{iL}^{M}=\delta\left(  X^{2}\right)  \left\{
\begin{array}
[c]{l}%
\frac{1}{2\sqrt{2}}F_{KL}^{a}X_{N}\left(  \Gamma^{KLN}\bar{\Gamma}^{M}%
-\eta^{NM}\Gamma^{KL}\right)  \lambda_{Li}^{a}\\
+\varepsilon_{ij}D_{K}\left(  X_{N}\varphi^{a}\right)  \left(  \Gamma
^{KN}\Gamma^{M}-\eta^{MN}\Gamma^{K}\right)  \lambda_{R}^{aj}\\
+D_{K}\left(  X_{N}\phi_{in}\right)  \left(  \Gamma^{KN}\Gamma^{M}-\eta
^{MN}\Gamma^{K}\right)  \eta_{R}^{n}\\
+\varepsilon_{ij}D_{K}\left(  X_{N}\phi^{\dagger jm}\right)  \left(
\Gamma^{KN}\Gamma^{M}-\eta^{MN}\Gamma^{K}\right)  \tilde{\eta}_{Rm}\\
-\frac{ig}{\sqrt{2}}X_{N}\Gamma^{MN}\left[
\begin{array}
[c]{c}%
\left(  if_{abc}\varphi^{\dagger b}\varphi^{c}+\phi^{\dagger j}t_{a}\phi
_{j}\right)  \lambda_{iL}^{a}\\
-2\phi^{\dagger j}t_{a}\phi_{i}\lambda_{jL}^{a}\\
-2\varphi_{a}\varepsilon_{ij}\left(  \phi^{\dagger j}t_{a}\right)  ^{n}%
\eta_{nL}\\
-2\varphi_{a}\left(  t_{a}\phi_{i}\right)  _{m}\tilde{\eta}_{L}^{m}%
\end{array}
\right]
\end{array}
\right\}  \label{n2current}%
\end{equation}
This $J_{iL}^{M}$ is a doublet of SU$\left(  2\right)  $ and vector $\otimes$
left-handed Weyl spinor of SO$\left(  4,2\right)  .$

These fermionic currents are conserved $\partial_{M}J_{iL}^{M}=0$ when we use
the equations of motion derived from the $N=2$ action given above. The general
variation of the action with respect to each field contains terms proportional
to both $\delta\left(  X^{2}\right)  $ as well as $\delta^{\prime}\left(
X^{2}\right)  $ (which arises from integration by parts). The equations that
emerge from the $\delta^{\prime}\left(  X^{2}\right)  $ terms are called
kinematic equations, while those emerging from the $\delta\left(
X^{2}\right)  $ term are called dynamical equations. The kinematic equations
can be solved easily, and they can be interpreted as the covariant version of
one of the three Sp$\left(  2,R\right)  $ constraints of the underlying
worldline theory (namely the $X\cdot P=0$ constraint). The dynamical equations
correspond to another Sp$\left(  2,R\right)  $ constraint ($P^{2}=0$
constraint) after being covariantized and modified by the interactions.
Finally, because of the delta functions, all equations listed below must be
taken at $X^{2}=0,$ which is the third Sp$\left(  2,R\right)  $ constraint. It
should be emphasized that all equations of motion follow from the action.

The following SU$\left(  2\right)  $ covariant $N=2$ equations are the
kinematic equations of motion%
\begin{gather}
X^{N}F_{NM}^{a}=\left(  X\cdot D+1\right)  \varphi^{a}=\left(  X\cdot
D+1\right)  \phi_{in}=0,\label{kin}\\
\left(  X\cdot D+2\right)  \lambda_{Li}^{a}=\left(  X\cdot D+2\right)
\eta_{nL}=\left(  X\cdot D+2\right)  \tilde{\eta}_{nR}=0,
\end{gather}
while the following SU$\left(  2\right)  $ covariant $N=2$ equations are the
dynamical equations of motion
\begin{equation}
\left(  D_{M}F^{MN}\right)  ^{a}-\left\{
\begin{array}
[c]{c}%
igf^{abc}X_{M}\overline{\lambda}_{L}^{bi}\Gamma^{MN}\lambda_{Li}^{c}%
+gf^{abc}\varphi^{\dagger b}\overleftrightarrow{D}^{N}\varphi^{c}\\
-gX_{M}\overline{\eta_{L}}^{n}\Gamma^{MN}t^{a}\eta_{nL}+gX_{M}\overline
{\tilde{\eta}_{R}}^{n}\Gamma^{MN}t^{a}\tilde{\eta}_{nR}+ig\phi^{\dagger
i}t^{a}\overleftrightarrow{D}^{M}\phi_{i}%
\end{array}
\right\}  =0, \label{dyn1}%
\end{equation}%
\begin{equation}
D^{2}\varphi^{a}+g^{2}f^{abc}f_{bde}\varphi^{c}\varphi^{\dagger d}\varphi
^{e}+\frac{ig}{\sqrt{2}}\varepsilon^{ij}f^{abc}\overline{\lambda_{Ri}}^{b}%
\bar{X}\lambda_{Lj}^{c}-g^{2}\left(  \phi^{\dagger i}\left\{  t^{a}%
,t^{b}\right\}  \phi_{i}\right)  \varphi_{b}^{\dagger}=0, \label{dyn2}%
\end{equation}%
\begin{equation}
D^{2}\phi^{\dagger in}+\left\{
\begin{array}
[c]{c}%
\sqrt{2}g\left(  t_{a}\right)  _{m}^{~n}\left(  \varepsilon^{ij}%
\overline{\tilde{\eta}_{R}}^{m}\bar{X}\lambda_{jL}^{a}+\overline{\eta_{L}}%
^{m}X\lambda_{R}^{aj}\right) \\
-g^{2}\left(  \phi^{\dagger i}t_{a}t_{b}\right)  ^{n}\left(  \varphi
^{a}\varphi^{\dagger b}+\varphi^{b}\varphi^{\dagger a}\right)  -g^{2}\left(
\phi_{j}^{\dagger}t_{a}\right)  ^{n}\phi^{\dagger(i}t_{a}\phi^{j)}%
\end{array}
\right\}  =0,
\end{equation}%
\begin{gather}
iX\bar{D}\lambda_{iL}^{a}+i\sqrt{2}g\varepsilon_{ij}f_{abc}\varphi^{b}%
X\lambda_{R}^{cj}-\sqrt{2}g\varepsilon_{ij}\left(  \phi^{\dagger j}%
t_{a}\right)  _{n}X\tilde{\eta}_{Rn}+\sqrt{2}g\left(  t_{a}\phi_{i}\right)
_{n}X\eta_{R}^{n}=0,\label{dyn3}\\
i\bar{X}D\lambda_{R}^{ai}+i\sqrt{2}g\varepsilon^{ij}f_{abc}\varphi^{\dagger
b}\bar{X}\lambda_{jL}^{c}-\sqrt{2}g\left(  \phi^{\dagger i}t_{a}\right)
_{n}\bar{X}\eta_{Ln}-\sqrt{2}g\varepsilon^{ij}\left(  t_{a}\phi_{j}\right)
_{n}\bar{X}\tilde{\eta}_{L}^{n}=0,\\
iX\bar{D}\eta_{nL}+\sqrt{2}g\left(  t_{a}\phi_{i}\right)  _{n}X\lambda
_{R}^{ai}+\sqrt{2}g\varphi^{\dagger a}X\left(  t_{a}\tilde{\eta}_{R}\right)
_{n}=0,\\
i\bar{X}D\tilde{\eta}_{nR}-\sqrt{2}g\left(  t_{a}\phi_{i}\right)
_{n}\varepsilon^{ij}\bar{X}\lambda_{jL}^{a}-\sqrt{2}g\varphi^{a}\bar{X}\left(
t_{a}\eta_{L}\right)  _{n}=0. \label{dyn7}%
\end{gather}

The $N=2$ SUSY transformations{} for the action associated with the
supercurrent in Eq.(\ref{n2current}) are (without auxiliary fields)%
\begin{equation}
\delta_{\varepsilon}A_{M}^{a}=-\frac{1}{\sqrt{2}}\overline{\varepsilon_{L}%
}^{i}\Gamma_{M}\bar{X}\lambda_{Li}^{a}+X^{2}\left[
\begin{array}
[c]{c}%
\frac{1}{2\sqrt{2}}~\overline{\varepsilon_{L}}^{i}\Gamma_{MN}D^{N}\lambda
_{Li}^{a}-\frac{g}{4}f^{abc}\varepsilon_{ij}\left(  \overline{\varepsilon}%
^{i}\Gamma_{M}\lambda_{R}^{bj}\right)  \varphi^{c}\\
-\frac{ig}{4}\left(  \overline{\eta_{L}}^{n}\Gamma_{M}\varepsilon_{R}%
^{i}\right)  \left(  t_{a}\phi_{i}\right)  _{n}-\varepsilon^{ij}\frac{ig}%
{4}\left(  \overline{\tilde{\eta}_{R}}^{n}\bar{\Gamma}_{M}\varepsilon
_{Li}\right)  \left(  t_{a}\phi_{j}\right)  _{n}%
\end{array}
\right]  +h.c. \label{n2susy1}%
\end{equation}%
\begin{equation}
\delta_{\varepsilon}\phi_{in}=\overline{\varepsilon_{R}}_{i}\bar{X}\eta
_{Ln}-\varepsilon_{ij}\overline{\varepsilon_{L}}^{j}X\tilde{\eta}_{Rn}%
+X^{2}\left[
\begin{array}
[c]{c}%
-\frac{1}{2}\overline{\varepsilon_{R}}_{i}\overline{D}\eta_{Ln}+\frac{1}%
{2}\varepsilon_{ij}\overline{\varepsilon_{L}}^{j}D\tilde{\eta}_{Rn}\\
\frac{ig}{\sqrt{2}}g\varphi_{a}^{\dagger}\overline{\varepsilon_{R}}_{i}\left(
t^{a}\tilde{\eta}_{R}\right)  _{n}+\frac{ig}{\sqrt{2}}\varepsilon_{ij}%
g\varphi_{a}\overline{\varepsilon_{L}}^{j}\left(  t^{a}\eta_{L}\right)  _{n}\\
+\frac{ig}{2\sqrt{2}}\left(  t_{a}\phi_{i}\right)  _{n}\left(  \overline
{\varepsilon_{L}}^{j}\lambda_{Lj}+\overline{\lambda_{L}}^{j}\varepsilon
_{Lj}\right) \\
-\frac{ig}{\sqrt{2}}\left(  t_{a}\phi_{j}\right)  _{n}\overline{\varepsilon
_{L}}^{j}\lambda_{Li}-\frac{ig}{\sqrt{2}}\left(  t_{a}\phi_{j}\right)
_{n}\overline{\lambda_{L}}^{j}\varepsilon_{Li}%
\end{array}
\right]  \label{n2susy4}%
\end{equation}%
\begin{equation}
\delta_{\varepsilon}\varphi^{a}=\varepsilon^{ij}\overline{\varepsilon_{R}}%
_{i}\bar{X}\lambda_{Lj}^{a}+X^{2}\left[
\begin{array}
[c]{c}%
-\frac{1}{2}\varepsilon^{ij}\overline{\varepsilon_{R}}_{i}\overline{D}%
\lambda_{Lj}^{a}-\frac{g}{2\sqrt{2}}f^{abc}\varphi^{b}\left(  \overline
{\lambda_{L}}^{i}\varepsilon_{Li}+\overline{\varepsilon_{L}}^{i}\lambda
_{Li}\right) \\
+i\frac{g}{\sqrt{2}}\phi^{\dagger ni}\overline{\varepsilon_{R}}_{i}\left(
t_{a}\tilde{\eta}_{R}\right)  _{n}-i\frac{g}{\sqrt{2}}\varepsilon
^{ij}\overline{\eta_{L}}^{n}\varepsilon_{Li}\left(  t_{a}\phi_{j}\right)  _{n}%
\end{array}
\right]  \label{n2susy3}%
\end{equation}%
\begin{equation}
\delta_{\varepsilon}\lambda_{Li}^{a}=\left\{
\begin{array}
[c]{c}%
-\varepsilon_{ij}i\left(  D_{M}\varphi\right)  ^{a}\left(  \gamma
^{M}\varepsilon_{R}^{j}\right)  +i\frac{1}{2\sqrt{2}}F_{MN}^{a}\left(
\gamma^{MN}\varepsilon_{Li}\right) \\
-i\frac{g}{\sqrt{2}}f^{abc}\varphi^{\dagger b}\varphi^{c}\varepsilon
_{Li}+\frac{g}{\sqrt{2}}\left[  2\varepsilon_{Lj}\phi^{\dagger j}t_{a}\phi
_{i}-\varepsilon_{Li}\phi^{\dagger j}t_{a}\phi_{j}\right]
\end{array}
\right\}  \label{n2susy2}%
\end{equation}%
\begin{align}
\delta_{\varepsilon}\eta_{Ln}  &  =i\left(  D_{M}\phi_{i}\right)  _{n}%
\Gamma^{M}\varepsilon_{R}^{i}+\varepsilon^{ij}\sqrt{2}g\left(  \varphi
^{\dagger}\phi_{j}\right)  _{n}\varepsilon_{Li}\label{n2susy5}\\
\delta_{\varepsilon}\tilde{\eta}_{Rn}  &  =i\varepsilon^{ij}\left(  D_{M}%
\phi_{i}\right)  _{n}\bar{\Gamma}^{M}\varepsilon_{Lj}+\sqrt{2}g\left(
\varphi\phi_{i}\right)  _{n}\varepsilon_{R}^{i} \label{n2susy6}%
\end{align}

The $N=2$ SUSY transformation above have some parallels to naive $N=2$ SUSY
transformations that one may attempt to write down as a direct generalization
from $3+1$ to $4+2$ dimensions. However, there are many features that are
completely different. Once we notice the parallels, part of the structure can
be understood from the spacetime SU$\left(  2,2\right)  $ group theory, as in
Eq.(\ref{su22rule}). The generalized features include the insertions that
involve $X$ $=X^{M}\Gamma_{M}$ or $\bar{X}$ $=X^{M}\bar{\Gamma}_{M},$ and the
terms proportional to $X^{2}.$ These are \textit{off-shell} SUSY
transformations that include interactions and leave the off-shell action invariant.

Despite all of the changes compared to naive SUSY, this SUSY symmetry provides
a representation of the supergroup SU$\left(  2,2|2\right)  $. This is
signaled by the fact that all terms are covariant under the bosonic subgroup
SU$\left(  2,2\right)  \otimes$SU$\left(  2\right)  ,$ while the complex
fermionic parameter $\varepsilon_{Li}$ and its conjugate $\overline
{\varepsilon_{L}}^{i}$ are in the $4,4^{\ast}$ representations of SU$\left(
2,2\right)  $, and are doublets of the R-symmetry SU$\left(  2\right)  ,$ as
would be expected for SU$\left(  2,2|2\right)  .$

The closure of the SUSY transformations is discussed for the case of $N=1$ in
Appendix (B) of reference \cite{susy2tN1}. The closure in that case was
SU$\left(  2,2|1\right)  $ when the fields are on-shell. It is straightforward
but tedious to verify that for the present case of $N=2,$ the closure is
SU$\left(  2,2|2\right)  $ \textit{when the fields are on-shell}. The SUSY
transformations above are actually off-shell. The closure off-shell goes
beyond SU$\left(  2,2|2\right)  $ and includes 2T-physics gauge
transformations (terms proportional to $X^{2}$ and other kinematic constraints
that do not vanish off-shell) of the type discussed in \cite{2tstandardM} and
\cite{susy2tN1}.

When reduced to $3+1$ dimensions, by solving the kinematic equations
(\ref{kin}) in a special gauge which we call the conformal
gauge\footnote{Dirac initiated a similar set of field equations on the
hypercone (without an action principle) to explain conformal symmetry
SO$\left(  4,2\right)  $ as the rotation group in 6 dimensions \cite{Dirac}%
-\cite{vasiliev}. A worldline approach along Dirac's ideas was also pursued
\cite{marnelius}-\cite{marnelius06}. From the point of view of 2T-physics,
Dirac's view of conformal symmetry amounts to only one of the shadows, which
we call the conformal shadow. The Sp$\left(  2,R\right)  $ phase space gauge
symmetry in 2T-physics, which was absent in previous work, was motivated by
signals of 2T in the supersymmetry structure of M theory \cite{Stheory}%
-\cite{sezgin} and it developed independently, unaware of Dirac's work. This Sp$\left(  2,R\right)  $ gauge
symmetry is at the root of the shadows and duality phenomena in 2T-physics. In the worldline theory the shadows
are obtained by making Sp$\left(  2,R\right)  $ gauge choices in phase space $\left(  X^{M},P_{M}\right)  $, while
in field theory the same shadows are recovered by solving the kinematic equations with various parameterizations
of spacetime as shown in \cite{emergentfieldth1}\cite{emergentfieldth2}.} described in
\cite{2tstandardM}-\cite{2tGravity}, the SU$\left(  2,2|2\right) $ transformations above reduce to a non-linear
off-shell realization of $N=2$ superconformal symmetry in $3+1$ dimensions.

\section{N=4 Super Yang-Mills in 4+2 dimensions}

The $N=4$ SYM multiplet has the same field content as the $N=2$ vector SYM
multiplet $\left(  A_{M}^{a},\lambda_{iL}^{a},\varphi^{a}\right)  $ coupled to
just one $N=2$ hypermultiplet $\left(  \phi_{i}^{a},\eta_{L}^{a},\eta_{R}%
^{a}\right)  $ whose fields are in the \textit{adjoint} representation. Thus,
all that we need to do is specialize the hypermultiplet in the previous
section to be in the adjoint representation labeled by $a$. Then there are
four left handed fermions ($\lambda_{iL}^{a},\eta_{L}^{a},\tilde{\eta}_{L}%
^{a}$) which we call $\lambda_{Lr}^{a}$ $r=1,2,3,4$ and six real scalars
(three complex ones, $\varphi^{a},$ $\phi_{i}^{a})$ which we call $\theta
_{u}^{a},$ $u=1,\cdots,6,$ in addition to the Yang-Mills field $A_{M}^{a},$
all in $4+2$ dimensions. In this section we present this structure directly in
an SU$(4)=$SO$\left(  6\right)  $ covariant way, thus displaying the $N=4$
SU$\left(  4\right)  $ $R$-symmetry. Then we show that the SU$\left(
4\right)  $ covariant theory agrees with the general form of the $N=1$ SUSY
theory of section 3, in four different rearrangements of the multiplets, thus
proving the $N=4$ SUSY symmetry in a different way.

Let $r,$ $s$ label the SU$(4)$ fundamental or antifundamental representations
(spinors of SO$\left(  6\right)  $) and let $u,v$ label the vector of
SO$\left(  6\right)  ,$ while $\left(  \alpha,\dot{\alpha}\right)  $ and $M$
label the spacetime SO$\left(  4,2\right)  $ spinor and vector representations
respectively. The manifestly SO$\left(  4,2\right)  \otimes$SU$\left(
4\right)  \otimes G$ invariant action can be written as $S^{N=4}=\int
d^{4+2}X~\delta\left(  X^{2}\right)  L^{N=4}\left(  X\right)  ,$ with%
\begin{equation}
L^{N=4}=\left\{
\begin{array}
[c]{c}%
-\frac{1}{4}F_{MN}^{a}F_{a}^{MN}+\frac{1}{2}\theta_{u}^{a}D^{M}D_{M}\theta
_{u}^{a}-\frac{g^{2}}{4}%
{\displaystyle\sum}
|f_{abc}\theta_{u}^{b}\theta_{v}^{c}|^{2}\\
+\frac{i}{2}\left[  \overline{\lambda_{L}}^{ar}X\bar{D}\lambda_{Lr}%
^{a}+gf_{abc}\left(  \lambda_{Lr}^{a}C\bar{X}\lambda_{Ls}^{b}\right)  \left(
\bar{\gamma}^{u}\right)  ^{rs}\theta_{u}^{c}\right]  +h.c.
\end{array}
\right\}  \label{n4action1}%
\end{equation}
Here $\gamma_{rs}^{u}=-\gamma_{sr}^{u}$ (and their Hermitian conjugates
$\left(  \bar{\gamma}^{u}\right)  ^{rs}$) are antisymmetric SO$\left(
6\right)  =$SU$\left(  4\right)  $ gamma matrices in a Weyl basis that satisfy
$\left(  \gamma^{u}\bar{\gamma}^{v}+\gamma^{v}\bar{\gamma}^{u}\right)
_{r}^{~s}=2\delta^{uv}\delta_{r}^{~s}.$ The explicit matrix form of the
antisymmetric SO$\left(  6\right)  $ gamma matrices $\left(  \gamma
^{u}\right)  _{rs},\left(  \bar{\gamma}_{u}\right)  ^{rs}$ can be taken as
\begin{equation}
\gamma^{u}=\left[  \left(  \sigma_{2}\times i\sigma_{2}\vec{\sigma}\right)
,\left(  \sigma_{2}\vec{\sigma}\times\sigma_{2}\right)  \right]
,\;\bar{\gamma}_{u}=\left[  \left(  \sigma_{2}\times i\sigma_{2}\vec{\sigma
}^{\ast}\right)  ,\left(  -\sigma_{2}\vec{\sigma}^{\ast}\times\sigma
_{2}\right)  \right]  \label{gammao62}%
\end{equation}
where $\bar{\gamma}_{u}$ is related to $\gamma^{u}$ by Hermitian conjugation
$\bar{\gamma}_{u}=\left(  \gamma^{u}\right)  ^{\dagger}$ or by complex
conjugation $\bar{\gamma}_{u}=-\left(  \gamma^{u}\right)  ^{\ast}$ (note
$\left(  -\sigma_{2}\vec{\sigma}^{\ast}\right)  =\left(  i\sigma
_{3},1,-i\sigma_{1}\right)  =\vec{\sigma}\sigma_{2}$). They satisfy the
property $\gamma_{rs}^{u}=$ $\frac{1}{2}\varepsilon_{rspq}\left(  \bar{\gamma
}^{u}\right)  ^{pq}.$ Using these one can recast the six independent real
scalar fields $\theta_{u}^{a}$ into an SU$\left(  4\right)  $ antisymmetric
tensor form $\left(  \varphi^{a}\right)  _{rs}$
\begin{equation}
\left(  \varphi^{a}\right)  _{rs}=\frac{1}{\sqrt{2}}\gamma_{rs}^{u}\theta
_{u}^{a}\text{ ~or~\ }\left(  \bar{\varphi}_{a}\right)  ^{rs}=\frac{1}%
{\sqrt{2}}\left(  \bar{\gamma}^{u}\right)  ^{rs}\theta_{u}^{a}=\left(
\varphi_{a}^{\dagger}\right)  ^{rs}=-\left(  \varphi_{rs}^{a}\right)  ^{\ast}%
\end{equation}
Because the complex conjugate is not independent there is a SU$\left(
4\right)  $ covariant duality relation
\begin{equation}
\left(  \bar{\varphi}^{a}\right)  ^{rs}=\frac{1}{2}\varepsilon^{rspq}%
\varphi_{pq}^{a}.
\end{equation}
This implies that the antisymmetric SU$\left(  4\right)  $ tensor $\left(
\varphi^{a}\right)  _{rs}$ contains only 3 independent complex numbers for
each $a,$ which is seen explicitly as follows
\begin{align}
\left(  \varphi^{a}\right)  _{rs}  &  =\frac{1}{\sqrt{2}}\left(
\begin{array}
[c]{cccc}%
0 & i\theta_{5}^{a}+\theta_{6}^{a} & \theta_{2}^{a}+i\theta_{3}^{a} &
-i\theta_{1}^{a}+\theta_{4}^{a}\\
-i\theta_{5}^{a}-\theta_{6}^{a} & 0 & -i\theta_{1}^{a}-\theta_{4}^{a} &
\theta_{2}^{a}-i\theta_{3}^{a}\\
-\theta_{2}^{a}-i\theta_{3}^{a} & i\theta_{1}^{a}+\theta_{4}^{a} & 0 &
i\theta_{5}^{a}-\theta_{6}^{a}\\
i\theta_{1}^{a}-\theta_{4}^{a} & -\theta_{2}^{a}+i\theta_{3}^{a} &
-i\theta_{5}^{a}+\theta_{6}^{a} & 0
\end{array}
\right) \\
&  \equiv\left(
\begin{array}
[c]{cccc}%
0 & -\bar{\varphi}^{3a} & \bar{\varphi}^{2a} & \varphi_{1}^{a}\\
\bar{\varphi}^{3a} & 0 & -\bar{\varphi}^{1a} & \varphi_{2}^{a}\\
-\bar{\varphi}^{2a} & \bar{\varphi}^{1a} & 0 & \varphi_{3}^{a}\\
-\varphi_{1}^{a} & -\varphi_{2}^{a} & -\varphi_{3}^{a} & 0
\end{array}
\right)  ,~\text{where }%
\genfrac{}{}{0pt}{}{\bar{\varphi}^{ia}\equiv\left(  \varphi_{i}^{a}\right)
^{\ast}}{i=1,2,3}
\label{su3scalars}%
\end{align}
This relation is useful to write the $N=4$ theory in an $N=1$ or $N=2$ basis.

\subsection{N=4 Super Yang-Mills as coupled N=1 supermultiplets}

We now want to verify that the SU$\left(  4\right)  $ covariant structures
above have $N=4$ supersymmetry and are in agreement with the $N=1$
supersymmetry structures in $4+2$ dimensions that we discussed in
Eqs.(\ref{N1action}-\ref{BF}). To do this we split the SU$\left(  4\right)  $
R-symmetry into SU$\left(  3\right)  \times$U$\left(  1\right)  $ and identify
the U$\left(  1\right)  $ as the R-symmetry associated with $N=1$
supersymmetry while the SU$\left(  3\right)  $ part is considered as an
internal symmetry acting on three $N=1$ chiral multiplets. Of course, there
are 4 different ways of splitting 4 into 3+1, each one of these corresponds to
the different $N=1$ supersymmetries within the $N=4$ theory. In each case, the
$N=4$ vector supermultiplet splits into one $N=1$ vector supermultiplet plus 3
chiral multiplets that transform into each other as a triplet of SU$\left(
3\right)  .$

To be specific let the $1$ in 3+1 correspond to the fourth member of the
SU$\left(  4\right)  $ quartet labeled as $r=\left(  i,4\right)  $ with
$i=1,2,3$. Then the SU$\left(  4\right)  $ quartet of fermions is split into a
SU$\left(  3\right)  $ triplet and a singlet $\lambda_{\alpha r}^{a}=\left(
\lambda_{\alpha i}^{a},\lambda_{\alpha4}^{a}\right)  .$ The singlet is
identified as the fermion in the $N=1$ vector multiplet $\left(  A_{M}%
^{a},\lambda_{\alpha}^{a}\right)  ,$ with $\lambda_{\alpha4}^{a}\equiv
\lambda_{\alpha}^{a}$ , while the triplet $\lambda_{\alpha i}^{a}$ belongs to
a $N=1$ chiral multiplet $\left(  \varphi_{i}^{a},\lambda_{\alpha i}%
^{a}\right)  ,$ with $r\rightarrow i=1,2,3$ labeling the fundamental
representation of SU$\left(  3\right)  .$ In this notation the kinetic term
for the fermions in the $N=4$ action is rewritten as%
\begin{equation}
\frac{i}{2}\overline{\lambda_{L}}^{ar}X\bar{D}\lambda_{Lr}^{a}+h.c.=\frac
{i}{2}\left[  \overline{\lambda_{L}}^{a}X\bar{D}\lambda_{aL}+\overline
{\lambda_{L}}^{ia}X\bar{D}\lambda_{Li}^{a}\right]  +h.c. \label{n4ton1_1}%
\end{equation}
We see that this is in agreement with the $N=1$ SUSY structure given in
Eq.(\ref{n11}), when the chiral multiplet $\left(  \varphi_{i}^{a}%
,\lambda_{\alpha i}^{a}\right)  $ is in the adjoint representation of the
gauge group $G$. Note that here SU$\left(  3\right)  $ with its label $i$ is a
global, not a local, symmetry.

Next we verify the same property for the scalars. The 3 complex scalars
$\varphi_{i}$ that appear in Eq.(\ref{su3scalars}) correspond to the 6 real
scalars $\theta_{u}^{a},$ $u=1,2,\cdots,6$ with the following identification%
\begin{align}
\left(  \varphi^{a}\right)  _{i4}  &  =\frac{1}{\sqrt{2}}\gamma_{i4}^{u}%
\theta_{u}^{a}=\varphi_{i}^{a},\;\;\left(  \varphi^{a}\right)  _{ij}=\frac
{1}{\sqrt{2}}\gamma_{ij}^{u}\theta_{u}^{a}=-\varepsilon_{ijk}\bar{\varphi
}^{ka},\\
\left(  \bar{\varphi}_{a}\right)  ^{i4}  &  =\frac{1}{\sqrt{2}}\left(
\bar{\gamma}^{u}\right)  ^{i4}\theta_{u}^{a}=-\bar{\varphi}^{ai},\;\;\left(
\bar{\varphi}_{a}\right)  ^{ij}=\frac{1}{\sqrt{2}}\left(  \bar{\gamma}%
^{u}\right)  ^{ij}\theta_{u}^{a}=\varepsilon^{ijk}\varphi_{k}^{a}
\label{n4ton1_2}%
\end{align}
Then the kinetic term for the scalars in the $N=4$ action is rewritten as%
\begin{equation}
\frac{1}{2}\theta_{a}^{u}D^{M}D_{M}\theta_{u}^{a}=\frac{1}{2}\bar{\varphi
}^{ia}D^{M}D_{M}\varphi_{i}^{a}+\frac{1}{2}\varphi_{i}^{a}D^{M}D_{M}%
\bar{\varphi}^{ia} \label{kinScalars}%
\end{equation}
This is in agreement with the $N=1$ SUSY structures in Eq.(\ref{n11}).
Furthermore, the Yukawa term in the $N=4$ action takes the form%
\begin{align}
&  \frac{i}{2}gf_{abc}\left(  \lambda_{Lr}^{a}C\bar{X}\lambda_{Ls}^{b}\right)
\left(  \bar{\gamma}^{u}\right)  ^{rs}\theta_{u}^{c}+h.c.\\
&  =\frac{i}{\sqrt{2}}gf_{abc}\left(  \lambda_{Lr}^{a}C\bar{X}\lambda_{Ls}%
^{b}\right)  \left(  \bar{\varphi}^{c}\right)  ^{rs}+h.c.\\
&  =\left[  \frac{i}{\sqrt{2}}2gf_{abc}\left(  \lambda_{Li}^{a}C\bar{X}%
\lambda_{L4}^{b}\right)  \left(  \bar{\varphi}^{c}\right)  ^{i4}+\frac
{i}{\sqrt{2}}gf_{abc}\left(  \lambda_{Li}^{a}C\bar{X}\lambda_{Lj}^{b}\right)
\left(  \bar{\varphi}^{c}\right)  ^{ij}\right]  +h.c.\\
&  =\left[  -i\sqrt{2}gf_{abc}\lambda_{Li}^{a}C\bar{X}\lambda_{L}^{b}%
\bar{\varphi}^{ic}+\frac{i}{\sqrt{2}}gf_{abc}\varepsilon^{ijk}\lambda_{Li}%
^{a}C\bar{X}\lambda_{Lj}^{b}\varphi_{k}^{c}\right]  +h.c. \label{yukawa}%
\end{align}
This is in agreement with the $N=1$ SUSY structures in Eq.(\ref{n12}) provided
the superpotential $W\left(  \varphi_{i}\right)  $ is
\begin{equation}
W\left(  \varphi\right)  =-\frac{g}{3\sqrt{2}}\varepsilon^{ijk}f_{abc}%
\varphi_{i}^{b}\varphi_{j}^{c}\varphi_{k}^{a}=-\sqrt{2}gf_{abc}\varphi_{1}%
^{b}\varphi_{2}^{c}\varphi_{3}^{a}, \label{W}%
\end{equation}

Next we rewrite the potential term $V\left(  \theta\right)  $ in the $N=4$
action in terms of the complex scalars $\varphi_{i}^{a}$ as follows
\begin{align}
V\left(  \theta\right)   &  =\frac{g^{2}}{4}%
{\displaystyle\sum}
|f_{abc}\theta_{u}^{b}\theta_{v}^{c}|^{2}=\frac{g^{2}}{4}f_{abc}f_{ab^{\prime
}c^{\prime}}\left(  \theta^{b}\cdot\theta^{b^{\prime}}\right)  \left(
\theta^{c}\cdot\theta^{c^{\prime}}\right) \\
&  =\frac{g^{2}}{4}f_{abc}f_{ab^{\prime}c^{\prime}}\left(  \bar{\varphi}%
^{ib}\varphi_{i}^{b^{\prime}}+\bar{\varphi}^{ib^{\prime}}\varphi_{i}%
^{b}\right)  \left(  \bar{\varphi}^{jc}\varphi_{j}^{c^{\prime}}+\bar{\varphi
}^{jc^{\prime}}\varphi_{j}^{c}\right) \\
&  =\frac{g^{2}}{2}\left(  f_{abc}\bar{\varphi}^{ib}\bar{\varphi}^{jc}\right)
\left(  f_{ab^{\prime}c^{\prime}}\varphi_{i}^{b^{\prime}}\varphi
_{j}^{c^{\prime}}\right)  +\frac{g^{2}}{2}\left(  f_{abc}\varphi_{i}^{b}%
\bar{\varphi}^{jc}\right)  \left(  f_{ab^{\prime}c^{\prime}}\bar{\varphi
}^{ib^{\prime}}\varphi_{j}^{c^{\prime}}\right)
\end{align}
The Jacobi identity%
\begin{equation}
f_{abc}f_{ab^{\prime}c^{\prime}}=f_{acb^{\prime}}f_{ac^{\prime}b}%
+f_{ab^{\prime}b}f_{ac^{\prime}c}%
\end{equation}
is used to rewrite the second term in the last line as%
\begin{equation}
\frac{g^{2}}{2}\left(  f_{abc}\varphi_{i}^{b}\bar{\varphi}^{jc}\right)
\left(  f_{ab^{\prime}c^{\prime}}\bar{\varphi}^{ib^{\prime}}\varphi
_{j}^{c^{\prime}}\right)  =\left[
\begin{array}
[c]{c}%
\frac{g^{2}}{2}\left(  f_{acb^{\prime}}\bar{\varphi}^{jc}\bar{\varphi
}^{ib^{\prime}}\right)  \left(  f_{ac^{\prime}b}\varphi_{j}^{c^{\prime}%
}\varphi_{i}^{b}\right) \\
+\frac{g^{2}}{2}\left(  f_{ab^{\prime}b}\bar{\varphi}^{ib^{\prime}}\varphi
_{i}^{b}\right)  \left(  f_{ac^{\prime}c}\varphi_{j}^{c^{\prime}}\bar{\varphi
}^{jc}\right)
\end{array}
\right]
\end{equation}
So the potential $V\left(  \theta\right)  =V\left(  \varphi\right)  $ takes
the form%
\begin{equation}
V\left(  \varphi\right)  =g^{2}\left(  f_{abc}\bar{\varphi}^{ib}\bar{\varphi
}^{jc}\right)  \left(  f_{ab^{\prime}c^{\prime}}\varphi_{i}^{b^{\prime}%
}\varphi_{j}^{c^{\prime}}\right)  -\frac{g^{2}}{2}\left(  f_{ab^{\prime}b}%
\bar{\varphi}^{ib^{\prime}}\varphi_{i}^{b}\right)  \left(  f_{acc^{\prime}%
}\bar{\varphi}^{jc}\varphi_{j}^{c^{\prime}}\right)
\end{equation}
We see that the potential can be written as the standard $N=1$ F and D terms
$V\left(  \varphi\right)  =V_{D}\left(  \varphi\right)  +V_{F}\left(
\varphi\right)  ,$
\begin{align}
V_{F}\left(  \varphi\right)   &  =\left(  gf_{abc}\bar{\varphi}^{ib}%
\bar{\varphi}^{jc}\right)  \left(  gf_{ab^{\prime}c^{\prime}}\varphi
_{i}^{b^{\prime}}\varphi_{j}^{c^{\prime}}\right)  =\bar{F}_{k}^{a}F_{a}%
^{k},\;\;\text{~\ with \ }\frac{1}{\sqrt{2}}F_{a}^{k}\varepsilon
_{kij}\text{\ }\equiv gf_{ab^{\prime}c^{\prime}}\varphi_{i}^{b^{\prime}%
}\varphi_{j}^{c^{\prime}}\\
V_{D}\left(  \varphi\right)   &  =\frac{1}{2}\left(  igf_{ab^{\prime}b}%
\bar{\varphi}^{ib^{\prime}}\varphi_{i}^{b}\right)  \left(  igf_{acc^{\prime}%
}\bar{\varphi}^{jc}\varphi_{j}^{c^{\prime}}\right)  =\frac{1}{2}B_{a}%
B^{a},\;\text{~\ with\ }B_{a}\equiv igf_{ab^{\prime}b}\bar{\varphi
}^{ib^{\prime}}\varphi_{i}^{b}%
\end{align}
This is in agreement again with the $N=1$ rules given in Eq.(\ref{n13}) when
the superpotential $W\left(  \varphi\right)  $ is precisely the one above in
Eq.(\ref{W}), since it then reproduces the correct $F$-term through $F_{a}%
^{k}=-\frac{\partial W}{\partial\varphi_{k}^{a}}.$

This agreement shows that the SU$\left(  4\right)  $ covariant theory has
$N=1$ SUSY in 4+2 dimensions for each of the four ways of reducing SU$\left(
4\right)  \rightarrow$SU$\left(  3\right)  \times$U$\left(  1\right)  .$ This
proves that the covariant theory has $N=4$ supersymmetry in 4+2 dimensions.
Indeed the evident SU$\left(  4\right)  $ $R$-symmetry implies that if there
is $N=1$ SUSY then there must be $N=4$ SUSY.

\subsection{N=4 covariant off-shell SUSY transformations in 4+2 dimensions}

Having established that the covariant action (\ref{n4action1}) has four
supersymmetries, it is useful to write the $N=4$ supersymmetry transformation
in covariant form as follows (using $\varepsilon_{R}^{s}\equiv C\overline
{\varepsilon_{Ls}}^{T}$, $\overline{\varepsilon_{R}}_{r}\equiv\left(
\varepsilon_{Lr}\right)  ^{T}C,$ and similarly for $\lambda_{R}^{as},\left(
\overline{\lambda_{R}^{a}}\right)  _{r}$)
\begin{equation}
\delta_{\varepsilon}A_{M}^{a}=\left\{  -\overline{\varepsilon_{L}}^{r}%
\Gamma_{M}\bar{X}\lambda_{Lr}^{a}+X^{2}\left[
\begin{array}
[c]{c}%
+\frac{1}{2}\overline{\varepsilon_{L}}^{r}\Gamma_{MN}D^{N}\lambda_{Lr}^{a}\\
+\frac{1}{4}gf_{abc}\left(  \overline{\varepsilon_{L}}^{r}\Gamma^{M}%
\lambda_{R}^{bs}\right)  \left(  \gamma^{u}\right)  _{rs}\theta_{u}^{c}%
\end{array}
\right]  \right\}  +h.c. \label{N4SUSYvector}%
\end{equation}%
\begin{equation}
\delta_{\varepsilon}\lambda_{Lr}^{a}=i\left(  D\theta_{u}^{a}\right)  \left(
\gamma^{u}\varepsilon_{R}\right)  _{r}+\frac{i}{2}F_{MN}^{a}\Gamma
^{MN}\varepsilon_{Lr}+\frac{i}{2}gf_{abc}\theta_{u}^{b}\theta_{v}^{c}\left(
\gamma^{uv}\varepsilon_{L}\right)  _{r} \label{N4SUSYspinor}%
\end{equation}%
\begin{equation}
\delta_{\varepsilon}\theta_{u}^{a}=\left\{  \left(  \bar{\gamma}_{u}\right)
^{rs}\overline{\varepsilon_{R}}_{r}\bar{X}\lambda_{Ls}^{a}+X^{2}\left[
-\frac{1}{2}\left(  \bar{\gamma}_{u}\right)  ^{rs}\overline{\varepsilon_{R}%
}_{r}\bar{D}\lambda_{Ls}^{a}+\frac{g}{2}f_{abc}\left(  \gamma_{uv}\right)
_{r}^{~s}\theta^{vb}\overline{\varepsilon_{L}}^{r}\lambda_{Ls}^{c}\right]
\right\}  +h.c. \label{N4SUSYscalar}%
\end{equation}
The first two expressions (\ref{N4SUSYvector},\ref{N4SUSYspinor}) may easily
be rewritten in terms of $\left(  \varphi^{a}\right)  _{rs}=\frac{1}{\sqrt{2}%
}\gamma_{rs}^{u}\theta_{u}^{a}.$ The last expression (\ref{N4SUSYscalar}) may
also be written in terms of $\left(  \varphi^{a}\right)  _{rs}$ as
follows\footnote{To compute the hermitian conjugate terms denoted as
\textquotedblleft$h.c.$\textquotedblright\ we recall from appendix A in
\cite{susy2tN1} the following rules which apply when all right handed fermions
are related to left handed fermions (Majorana fermions) as explained in
section (\ref{notation})%
\[%
\begin{array}
[c]{l}%
\left(  \overline{\psi_{1L}}\psi_{2L}\right)  ^{\dagger}=-\overline{\psi_{2L}%
}\psi_{1L}=\overline{\psi_{1R}}\psi_{2R},\\
\left(  \overline{\psi_{1L}}\Gamma^{M}\psi_{2R}\right)  ^{\dagger}%
=\overline{\psi_{2R}}\bar{\Gamma}^{M}\psi_{1L}=\overline{\psi_{1R}}\bar
{\Gamma}^{M}\psi_{2L},\\
\left(  \overline{\psi_{1L}}\Gamma^{M}\bar{\Gamma}^{N}\psi_{2L}\right)
^{\dagger}=-\overline{\psi_{2L}}\Gamma^{N}\bar{\Gamma}^{M}\psi_{1L}%
=\overline{\psi_{1R}}\Gamma^{M}\bar{\Gamma}^{N}\psi_{2R},\\
\left(  \overline{\psi_{1L}}\Gamma^{MN}\psi_{2L}\right)  ^{\dagger}%
=\overline{\psi_{2L}}\Gamma^{MN}\psi_{1L}=\overline{\psi_{1R}}\Gamma^{MN}%
\psi_{2R}%
\end{array}
\]
}
\begin{equation}
\left(  \delta\varphi^{a}\right)  _{rs}=\left\{
\begin{array}
[c]{c}%
-\sqrt{2}\left(  \overline{\varepsilon_{R}}_{r}\bar{X}\lambda_{Ls}%
^{a}-\overline{\varepsilon_{R}}_{s}\bar{X}\lambda_{Lr}^{a}\right)  +\sqrt
{2}\varepsilon_{rspq}\overline{\varepsilon_{L}}^{p}X\lambda_{R}^{aq}\\
+X^{2}\left[
\begin{array}
[c]{c}%
\frac{1}{\sqrt{2}}\left(  \overline{\varepsilon_{R}}_{r}\bar{D}\lambda
_{Ls}^{a}-\overline{\varepsilon_{R}}_{s}\bar{D}\lambda_{Lr}^{a}\right)
-\frac{1}{\sqrt{2}}\varepsilon_{rspq}\overline{\varepsilon_{L}}^{p}%
D\lambda_{R}^{aq}\\
+\frac{1}{2}gf_{abc}\left[  \left(  \overline{\varepsilon_{L}}\varphi
^{b}\right)  _{r}\lambda_{Ls}^{c}-\left(  \overline{\varepsilon_{L}}%
\varphi^{b}\right)  _{s}\lambda_{Lr}^{c}-\varepsilon_{rspq}\overline
{\varepsilon_{L}}^{p}\left(  \bar{\varphi}^{b}\lambda_{L}^{c}\right)
^{q}\right] \\
+\frac{1}{2}gf_{abc}\left[  \varepsilon_{rspq}\left(  \overline{\varepsilon
_{R}}\bar{\varphi}^{b}\right)  ^{p}\lambda_{R}^{cq}-\overline{\varepsilon_{R}%
}_{r}\left(  \varphi^{b}\lambda_{R}^{c}\right)  _{s}+\overline{\varepsilon
_{R}}_{s}\left(  \varphi^{b}\lambda_{R}^{c}\right)  _{r}\right]
\end{array}
\right]
\end{array}
\right\}  \label{N4SUSYscalar2}%
\end{equation}
To verify this last form we reconstruct $\delta\theta_{u}^{a}=-\frac{1}%
{2\sqrt{2}}\left(  \delta\varphi^{a}\right)  _{rs}\left(  \bar{\gamma}%
_{u}\right)  ^{rs}$ by inserting the expression in (\ref{N4SUSYscalar2}) and
obtain the expression in (\ref{N4SUSYscalar})\footnote{The following
properties of the SO$\left(  6\right)  $ gamma matrices are also useful:
$Tr\left(  \gamma_{u}\bar{\gamma}_{v}\right)  =4\delta_{uv}$ and $\left(
\gamma_{u}\right)  _{rs}\left(  \bar{\gamma}^{u}\right)  ^{pq}=-2\left(
\delta_{r}^{p}\delta_{s}^{q}-\delta_{s}^{p}\delta_{r}^{q}\right)  $ and
$\left(  \gamma_{u}\right)  _{rs}\left(  \gamma^{u}\right)  _{pq}%
=-2\varepsilon_{rspq}.$}.

The $N=4$ SUSY transformations (\ref{N4SUSYvector}-\ref{N4SUSYscalar2}) are
obtained by SU$\left(  4\right)  $ covariantizing the $N=1$ transformations
given in \cite{susy2tN1} (for comparison we define $\varepsilon_{L4}%
=\varepsilon_{L}/\sqrt{2}$). The $N=1$ SUSY formulas combined with SU$\left(
4\right)  $ insure that they work for $N=4$ SUSY.

Furthermore, by rewriting the $N=4$ transformations in the $N=2$ basis, it can
be verified that they are also in agreement with the $N=2$ transformations in
Eqs.(\ref{n2susy1}-\ref{n2susy6}) by using the following identification of
$N=4$ and $N=2$ degrees of freedom%
\begin{equation}
\lambda_{rL}^{a}=\left(
\begin{array}
[c]{c}%
\lambda_{iL}^{a}\\
\eta_{L}^{a}\\
-\tilde{\eta}_{L}^{a}%
\end{array}
\right)  ,\text{ }\varepsilon_{rL}=\left(
\begin{array}
[c]{c}%
\frac{1}{\sqrt{2}}\varepsilon_{iL}\\
0\\
0
\end{array}
\right)  ,\text{ }i=1,2 \label{N4N2spinorIdent}%
\end{equation}%
\begin{equation}%
\begin{array}
[c]{l}%
\varphi_{i3}^{a}=-\phi_{i}^{a},\text{ }\varphi_{i4}^{a}=\varepsilon_{ij}%
\bar{\phi}^{aj},\text{ }\varphi_{34}^{a}=\bar{\varphi}^{a},\text{ }%
\varphi_{ij}^{a}=-\varepsilon_{ij}\varphi^{a}\\
\bar{\varphi}^{ai3}=\bar{\phi}^{ia},\text{ }\bar{\varphi}^{ai4}=-\varepsilon
^{ij}\phi_{j}^{a},\text{ }\bar{\varphi}^{a34}=-\varphi^{a},\text{ }%
\bar{\varphi}^{aij}=\varepsilon^{ij}\bar{\varphi}^{a}%
\end{array}
\label{N4N2scalarIdent}%
\end{equation}

We emphasize that the off-shell SUSY transformations in 4+2 dimensions include
terms proportional to $X^{2}$ which are new structures as compared to SUSY
transformations in $3+1$ dimensions. The closure of these transformations
(commutators) is consistent with SU$\left(  2,2|4\right)  $ when the fields
are on-shell, but off shell there are additional terms beyond SU$\left(
2,2|4\right)  .$ The extra terms in the closure correspond to gauge
transformations that are the 2T gauge symmetries of 2T field theory of the
type discussed in \cite{2tstandardM}, and they are expected to vanish in the
gauge invariant sector of the theory.

The $N=4$ supercurrents associated with these SUSY transformations take the
form
\begin{equation}
J_{Lr}^{M}=\delta\left(  X^{2}\right)  \left\{
\begin{array}
[c]{c}%
\frac{1}{2}F_{PQ}^{a}X_{N}\left(  \Gamma^{PQN}\bar{\Gamma}^{M}-\eta^{NM}%
\Gamma^{PQ}\right)  \lambda_{Lr}^{a}\\
-\sqrt{2}\left(  \Gamma^{QP}\Gamma^{M}-\eta^{MP}\Gamma^{Q}\right)  \left[
D_{Q}\left(  X_{P}\varphi\right)  ^{a}\lambda_{R}^{a}\right]  _{r}~\\
-gf_{abc}\Gamma^{MN}X_{N}\left(  \varphi^{a}\bar{\varphi}^{b}\lambda_{L}%
^{c}\right)  _{r}%
\end{array}
\right\}
\end{equation}
Its expression in terms of $\theta_{u}^{a}$ is obtained by substituting
$\left(  \varphi^{a}\right)  _{rs}=\frac{1}{\sqrt{2}}\gamma_{rs}^{u}\theta
_{u}^{a}$. The $N=4$ supercurrents are conserved $\partial_{M}J_{rL}^{M}=0$
when the equations of motion that follow from the $N=4$ action are used. It
should be noted that the expression for the supercurrents can be modified by
terms of the form%
\begin{equation}
\Delta J_{Lr}^{M}=\delta\left(  X^{2}\right)  X^{M}\xi_{Lr}%
\end{equation}
that are automatically conserved $\partial_{M}\left(  \Delta J_{Lr}%
^{M}\right)  =0,$ when the spinors $\xi_{Lr}$ are arbitrary except for
satisfying the following homogeneity condition%
\begin{equation}
\left(  X\cdot\partial+4\right)  \xi_{Lr}=0,\;\text{equivalently\ }\xi
_{Lr}\left(  tX\right)  =t^{-4}\xi_{Lr}\left(  X\right)  .
\end{equation}
The currents $J_{Lr}^{M}$ above agree with the $N=1$ supercurrent in
\cite{susy2tN1} after inserting the $N=1$ basis discussed in
Eqs.(\ref{n4ton1_1}-\ref{n4ton1_2}) (for comparison with \cite{susy2tN1} we
define $\left(  J_{4L}^{M}\right)  _{N=4}\equiv\sqrt{2}\left(  J_{L}%
^{M}\right)  _{N=1}$ ). Furthermore, after inserting the $N=2$ basis of
Eqs.(\ref{N4N2spinorIdent},\ref{N4N2scalarIdent}), the $N=4$ currents above
also agree with the $N=2$ currents in Eq.(\ref{n2current}) when we identify
two of the $N=4$ currents with the $N=2$ currents up to a $\sqrt{2}$
normalization, $\left(  J_{iL}^{M}\right)  _{N=4}=\sqrt{2}\left(  J_{iL}%
^{M}\right)  _{N=2}$, $i=1,2.$ Of course, both $N=2$ and $N=4$ currents are
consistent with $N=1$.

\section{Physics consequences and future directions \label{conclude}}

In this paper we have explicitly constructed $N=2$ and $N=4$ supersymmetric
field theories in the theoretical framework of 2T-physics with two times. All
fields, including vectors and spinors, are in $4+2$ dimensional flat spacetime
that has a natural SO$(4,2)=$SU$\left(  2,2\right)  $ rotation symmetry, but
no translation symmetry. Although naively extra time dimensions lead to
troublesome negative norm ghosts, our theories are physical because they
include special gauge symmetries and kinematical constraints that insure
ghost-free unitary theories.

After gauge fixing and solving the kinematic constraints, our theories produce
conformal shadows in $3+1$ flat dimensions in which SO$(4,2)$ is the usual
conformal group that includes Poincar\'{e} symmetry, and hence translation
symmetry in $3+1$ dimensional Minkowski spacetime. These conformal shadows
coincide with previously established $N=1,2,4$ supersymmetric massless
renormalizable field theories with special forms of the superpotential.

In particular the famous $N=4$ super Yang-Mills theory in $3+1$ dimensions,
that continues to attract a lot of interest, is now seen to have a parent
theory in $4+2$ dimensions, that naturally explains its exact conformal
symmetry, and possibly some of its other properties as well.

An important aspect of 2T-physics is that it also produces many other shadows
in $3+1$ dimensions as explained in \cite{2treviews},\cite{2tHandAdS}%
-\cite{twistorBP} in the worldline context and in \cite{emergentfieldth1}%
\cite{emergentfieldth2} in the field theory context. By using the approach of
\cite{emergentfieldth1}\cite{emergentfieldth2} we can produce in a
straightforward way other dual shadows of our $N=1,2,4$ supersymmetric
theories in various curved spacetimes, including Robertson-Walker, AdS$_{4},$
dS$_{4},$ AdS$_{3}\times$S$^{1},$ AdS$_{2}\times$S$^{2},$ any maximally
symmetric spacetime, any conformally flat spacetime, some singular spacetimes,
all in $3+1$ dimensions. All of these share the full SO$\left(  4,2\right)  $
symmetry, \textit{as well as the full }SU$\left(  2,2|N\right)  $\textit{
supersymmetry} of the parent theory, realized in different forms as a hidden
symmetry in various spacetimes.

We expect that more shadows, that contain mass parameters as seen in the
worldline theories, can also be obtained in field theory, thus arriving at
very unusual realizations of the SU$\left(  2,2|N\right)  $ symmetry. All
shadows can be transformed into each other by the underlying Sp$\left(
2,R\right)  $ gauge symmetry which now plays the role of duality
transformations in field theory \cite{emergentfieldth1}\cite{emergentfieldth2}%
. It is expected that such duality properties of our theories can be used to
explore non-perturbative or exact solutions of $N=1,2,4$ supersymmetric
Yang-Mills theories.

In particular one may now revisit previous studies of supersymmetric theories,
including classical solutions, monopoles, instantons, Seiberg-Witten analysis
\cite{Seiberg-Witten}, $N=4$ dualities, AdS-CFT \cite{AdSCFT}, etc., but now
from the perspective of $4+2$ dimensions and using new tools in the context of
2T-physics. These will be explored in the future.

As in the case of the non-supersymmetric Standard Model in $4+2$ dimensions
\cite{2tstandardM}, we expect that the supersymmetric version produces a
shadow that includes certain constraints on the structure of the field theory
in $3+1$ dimensions that are not present in the usual approach in 1T field
theory. In particular generating masses for the fields is not as
straightforward as the ordinary 1T approach, and it requires the coupling of
the dilaton and hence of supergravity in $d+2$ dimensions (see footnote
(\ref{cubic})). At this point gravity in 2T field theory has been constructed
in $d+2$ dimensions \cite{2tGravity}. One of our future goals is to
supersymmetrize it and couple it to the $N=1,2,4$ theories constructed in this
paper. It is expected that the resulting structures will provide a number of
constraints on SUSY theories that could be of phenomenological interest in
case the LHC discovers supersymmetry.

Another future direction is SUSY theories in $d+2$ dimensions with $d\neq4.$
We remind the reader that $N=4$ super Yang-Mills theory in $3+1$ dimensions is
a reduced version of $N=1$ super Yang-Mills theory in $9+1$ dimensions.
Therefore, from the point of view of 2T-physics, it is natural to expect that
there must exist a SYM theory in $10+2$ dimensions which can be compactified
to our $N=4$ SYM theory in $4+2$ dimensions presented in this paper. Such a
theory breaks the 11-dimensional barrier for SUSY, but becomes physical with
the extra gauge symmetries and constraints supplied by 2T-physics. This will
be discussed in the near future in a paper on supersymmetric theory in higher
dimensions which is currently under preparation.

\begin{acknowledgments}
We gratefully acknowledge discussions with S-H. Chen, B. Orcal, and G. Quelin.
\end{acknowledgments}

\end{document}